\documentclass[a4paper,11pt]{article}
\usepackage[utf8]{inputenc}
\usepackage[english]{babel}

\usepackage{mathtools}
\usepackage{amsmath}
\usepackage{mathtools}
\usepackage[new]{old-arrows}
\usepackage{tikz-cd}

\usepackage{makecell}

\usepackage{amsthm}
\usepackage{amsfonts}
\usepackage[section]{placeins}

\usepackage[font={small,it}]{caption}

\usepackage{float}
\usepackage{hyperref}

\usepackage{tikz}
\usetikzlibrary{matrix,arrows,decorations}
\newtheorem{theorem}{Theorem}[section]

\theoremstyle{remark}
\newtheorem{remark}[theorem]{Remark}
\theoremstyle{definition}
\newtheorem{definition}[theorem]{Definition}

\title{Dynamic Initial Margin via Chebyshev Tensors}

\author{Mariano Zeron\footnote{m.zeron@mocaxintelligence.com}   \\ Ignacio Ruiz\footnote{i.ruiz@mocaxintelligence.com}}

\begin{document}
\maketitle

\begin{abstract}

We present two methods, based on Chebyshev tensors, to compute dynamic sensitivities of financial instruments within a Monte Carlo simulation. These methods are implemented and run in a Monte Carlo engine to compute Dynamic Initial Margin as defined by ISDA (SIMM). We show that the levels of accuracy, speed and implementation efforts obtained, compared to the benchmark (DIM obtained calling pricing functions such as are found in risk engines), are better than those obtained by alternative methods presented in the literature, such as regressions (\cite{Zhu Chan}) and Deep Neural Nets (\cite{DNNs IM}).


\end{abstract}

\section{Introduction}\label{sec: Intro}

One of the consequences of the $2008$ financial crisis has been a worldwide push for strong collateralisation of OTC derivative transactions. According to ISDA, the amount of global collateral posted as margin up to September $2017$ is of $\$1.41$ trillion U.S. dollars. The following table summarises the data breaking down margin into Variation Margin (VM) and Initial Margin (IM).\footnote{By now, the numbers must be much higher.}

\begin{table}[H]
\centering
\begin{tabular}{|l|l|l|l|}
\hline
 & Cleared ($\$$bn) & Uncleared ($\$$bn) & total ($\$$bn)\\
\hline
Variational Margin & $260.8$ & $870.4$	& $1,131.2$\\
Initial Margin & $173.4$ & $107.1$ & $280.5$\\
Total & $434.2$ & $977.5$ & $1,411.7$ \\
\hline
\end{tabular}
\caption{Cleared and Uncleared Variational and Initial Margin up to September $2017$}
\label{tab: Margin amounts in market}
\end{table}

As institutions and bilateral portfolios are migrated into the margining regime, the margin of uncleared derivatives should show the highest growth rate. It is expected that IM should reach the trillion range in a few years. 

Initial margin requirements translate into funding cost (funding rate for cash and repo rate for bonds), liquidity risk, and capital costs. Hence it is central for the profitability of financial institutions to manage these costs and risks, not only today but also in the future. Therefore, to optimise IM one must have a forward-looking view. This requires simulating Initial Margin inside a Monte Carlo (MC) simulation. We call simulated Initial Margin, \emph{Dynamic Initial Margin} (DIM).

Specific uses of a forward IM model include trade pricing (MVA), regulatory capital (IMM and CVA-FRTB), risk management (tail risk), stress testing and most likely, in the near future, accounting MVA. Hence, sound models for IM inside Monte Carlo simulations will be central for financial institutions going forward.

Initially, IM was defined as a $99\%$ percentile VaR computed using a risk horizon of $10$-days (\cite{Margin requirements}). However, qualified counterparties are allowed to use internal models to calculate IM. To simplify reconciliation between counterparties, the industry has adopted the Standard Initial Margin Model (SIMM) as the standard to compute IM. This is a model that replicates a $99\%$ VaR value using the sensitivities of the portfolio to specific risk factors, weighed appropriately by parameters calibrated during periods of stress. The specification of such calculation can be found in \cite{ISDA SIMM}.

Dynamic Initial Margin (DIM) is typically estimated using Monte Carlos simulations. For comparison purposes throughout the paper, we consider a Monte Carlo simulation consisting of $10,000$ paths and $100$ time points in the future. Computing a full set of PVs in this simulation (requirement in many risk calculations) has a computational cost or complexity of $O(10^6)$. In the case of DIM, assuming pricing functions are used, the computational cost is substantial. For the quantile based IM, where quantiles are computed over PnL distributions consisting of hundreds of values at each node of the simulation, the computational cost has order $O(10^8)$. In the case of SIMM, assuming an average of $10$-$50$ sensitivities per trade, the cost has order $O(10^7)$. Both these costs are prohibitively high; more so considering the already large number of risk calculations that need to be done on a regular basis.

The substantial computational load associated to the computation of DIM has forced practitioners in the industry to look for alternative ways of modelling it. Any useful approximating technique should have the following three requirements. It must be accurate; the numbers provided by the model should be reliable. It must be efficient; if a calculation needed on an hourly basis takes one day to compute, it is of no use in practice. Finally, it must be easy to implement and maintain; ideally, implementation must be modular, with tangible benefits in monthly time-frames, and easily done on existing platforms.

One of the simplest ways of estimating DIM is with regressions. Many papers deal with these techniques within this context (for example, \cite{Zhu Chan}, \cite{Pykhtin}). Generally, these estimate IM defined as the quantile of a distribution of PnLs. Their main advantage is the simplicity and speed of the regressions once they have been trained. The main drawbacks are the lack of accuracy, difficulty when used to estimate IM as defined by ISDA (SIMM), set of assumptions it makes (for example, normality of the PnL distribution) and the requirement of a full set of present values (PVs) in the Monte Carlo simulation; that is, a PV at each node of the simulation, which in itself imposes a pricing cost of $O(10^6)$.

Adjoint Algorithmic Differentiation (AAD), a technique familiar to most in the industry, has the advantage of computing sensitivities to a high level of precision. Moreover, it is unaffected in terms of precision and computational cost, by the number of sensitivities to compute. This makes it an ideal option for computing Dynamic SIMM to a high level of accuracy. However, it comes with the downside of a high computational cost, generally estimated to be between $5$ and $10$ times that of a typical CVA calculation (between $O(10^6)$ and $O(10^7)$, assuming, as above, a cost of $O(10^6)$ for a typical CVA calculation). Moreover, it comes with considerable implementation challenges. These challenges are often prohibitive when AAD is to be incorporated into already existing systems.

A recent technique, which has gathered a lot of attention, comes from Machine Learning. Namely Deep Neural Networks (see \cite{DNNs IM}). Neural Networks have the ability to approximate functions using relatively basic mathematical objects (in this case neurons). Once trained, they can be evaluated very efficiently. One of the main challenges with Neural Nets, and more generally Machine Learning algorithms, is hyper-parameter optimisation. This is a task that mainly relies on intuition and heuristic methods that aim at finding the right balance between under-fitting and over-fitting. In some cases, the hyper-parameters needed are relatively simple to find; in some others they are much more complicated and finding the right combination can be cumbersome.

Finally, the most relevant DIM computing technique for this paper, is based on Chebyshev tensors. As function approximators, Chebyshev tensors enjoy strong convergence properties. Moreover, once built, they are evaluated very efficiently. This means Chebyshev tensors accelerate risk calculations while maintaining high levels of accuracy. Chebyshev tensors can be used in a wide variety of risk calculations. For example, they can be used to estimate all sorts of CCR risk metrics such as CVA, exposure profiles and capital values under IMM (see \cite{MZ IR disclosure}). They can also be used with great success in Market Risk for calculations such as capital under FRTB-IMA (\cite{MZ IR FRTB}). In this paper, we take advantage of the strong mathematical properties enjoyed by Chebyshev tensors to compute trade sensitivities which are then used to estimate SIMM. We show substantial computational reductions obtained compared to the brute force approach, while keeping very high levels of accuracy.

The  paper is organised as follows. In Section \ref{sec: cheb tensors} we introduce Chebyshev tensors and the theory that supports the use these objects in all sorts of risk calculations. Section \ref{sec: Sensitivities with Cheb} presents how to use Chebyshev tensors to compute dynamic sensitivities, which in turn are used to compute DIM. Section \ref{sec: Results} presents the results obtained from the simulation of future IM (SIMM) using Chebyshev tensors. We also present the results of simulating DIM using regressions as described in \cite{Zhu Chan}. In Section \ref{sec: Results analysis} the accuracy and speed of these two methodologies are compared to the benchmark methodology; the benchmark being when the original pricing function is used to compute the partial derivatives at each node of the simulation using finite difference in a “brute-force” fashion. This Section also discusses the advantages and disadvantages of Chebyshev tensors in the computation of DIM with respect to other techniques, such as AAD and Deep Neural Nets.\footnote{These techniques are not tested in this paper.} A short conclusion is drawn in Section \ref{sec: Conclusion}.

\section{Chebyshev tensors and interpolants}\label{sec: cheb tensors}

Chebyshev tensors and Chebyshev interpolants lie at the heart of the techniques used to compute the results shown in Section \ref{sec: Results}. In this Section we cover the main definitions and mathematical properties that make Chebyhsev tensors such good function approximators. For further details on the theory of Chebyshev approximation we refer the reader to \cite{TrefethenTextbook},  and \cite{MoCaXChebUltra}.

\subsection{One-dimensional case}\label{sec: cheb interpolants dim one}

Polynomial interpolants are often thought as poor approximators. The bad reputation is owed in part to results that have been around for many decades. The first one is due to Runge who gave an example of an analytic function for which equidistant interpolation diverges exponentially \cite{Runge}. Analytic functions, by definition, enjoy a high degree of smoothness. Equidistant points are a natural choice for interpolation if there is no a-priori information to say otherwise. This example shows how polynomial interpolation, if not done properly, can have terrible consequences even on well behaved functions. The second result, which also goes back a long way, says that there is no interpolation scheme that guarantees convergence for the set of continuous functions \cite{Faber}.

Results such as the ones mentioned above cemented a belief that using polynomial interpolants as approximators of functions (even analytic ones) is not appropriate. Even textbooks in the subject of function approximation warn against the use of polynomial interpolants (Appendix in \cite{TrefethenTextbook}). What is often missed, is that interpolation on carefully chosen distribution of points can yield optimal approximation properties if applied to the correct class of functions.

\begin{definition}
The Chebyshev points associated with the natural number $n$ are the real part of the points
\end{definition}

\begin{equation*}
x_j = \mathrm{Re}(z_j) = \frac{1}{2}(z_j+z_j^{-1} ),\ \ \ \ \ 0\leq j \leq n.
\end{equation*}

Equivalently, Chebyshev points can be defined as

\begin{equation*}
x_j = \mathrm{cos} \Big( \frac{j\pi}{n} \Big), \ \ \ \ \ \ 0\leq j \leq n.  
\end{equation*}

These points are the result of projecting equidistant points on the upper half of the unitary circle onto the real line.

\begin{figure}[H]
\centering
\includegraphics[scale=0.5]{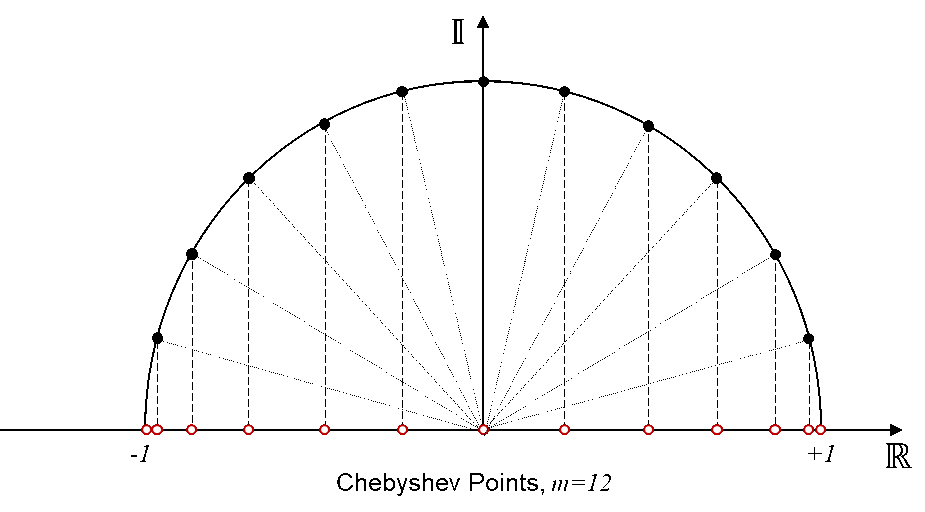}
\caption{Chebyshev points in one dimension.}
\end{figure}

The definition of Chebyshev points is given for an interval $[-1,1]$. This, however, can be extended to any interval $[a,b]$ by mapping $[-1,1]$ to $[a,b]$, with the aid of a linear transformation followed by a translation. Therefore, although most results in this Section are stated for functions defined on $[-1,1]$, or $[-1,1]^d$ in the case of $d$-dimensional functions, these are valid for more general domains $[a,b]$ and hyper-rectangles in higher-dimensions.

A set of points $x_0, \ldots, x_n$ along with a set of associated real values $v_0, \ldots, v_n$ constitute a \emph{tensor}. When the points $x_0, \ldots, x_n$ are Chebyshev points, then we have a \emph{Chebyshev tensor}. Polynomial interpolants are closely associated to tensors. Given a tensor with points $x_0, \ldots, x_n$ and values, $v_0, \ldots, v_n$, it is well known that there is a unique polynomial $p_n$ of order at most $n$ that interpolates the values $v_0, \ldots, v_n$ at the points $x_0, \ldots, x_n$. When the points $x_0, \ldots, x_n$ are Chebyshev points, we have a \emph{Chebyshev interpolant}. From now on, whenever Chebyhsev tensors are mentioned, we also refer to the unique Chebyshev polynomial they define.

One-dimensional Chebyshev tensors have unique convergence properties. 

\begin{theorem}\label{thm: convergence Lip}
Let $f$ be a Lipschitz continuous function on the interval $[-1,1]$. Then $p_n$ converges uniformly to $f$.
\end{theorem}
\noindent where $p_n$ is the Chebyshev interpolant to $f$ on the first $n+1$ Chebyshev points. We remind the reader that a function $f$ is Lipschitz continuous if given $\varepsilon>0$, there is a $K>0$ such that 
\begin{equation*}
|f(x) - f(y)|\leq K|x - y|
\end{equation*}
\noindent for all $x, y$ in the domain of $f$.

Theorem \ref{thm: convergence Lip} can be significantly strengthened by restricting the class of functions over which we work.

\begin{theorem}\label{thm: exponential convergence}
Let $f$ be an analytic function on the interval $[-1,1]$. Consider its analytical continuation to the open Bernstein ellipse $E_p$ of radius $\rho$, where it satisfies $|f(x)| \leq M$, for some $M$. Then for each $n\geq 0$
\begin{equation*}
\|f-p_n \|_{\infty}\leq \frac{4M\rho^{-n}}{\rho - 1}  
\end{equation*}
\end{theorem}

We remind the reader that a function $f$ is analytic if for all $x$ in the domain of $f$, the Taylor expansion at $x$ converges to to $f(x)$.

Theorem \ref{thm: exponential convergence} says that very few interpolation points are needed to get a high degree of accuracy when the function is analytic. In finance, most functions we deal with on a regular basis (pricing functions, sensitivity functions, etc) are analytic (at least piece-wise analytic). The use of Chebyshev tensors  and their corresponding interpolants therefore makes sense.

Building a Chebyshev interpolant for a function $f$ only requires specifying Chebyshev points $x_0, \ldots, x_n$ and obtaining their associated values $f(x_0), \ldots, f(x_n)$. To evaluate such interpolant, however, one must be careful. It is always important to make sure one works with robust and fast algorithms. It is often forgotten that rounding off errors on a computer can accumulate to the point that the values returned are completely different to what they should be in theory. A good example is one of the most popular algorithms for polynomial interpolation via Vandermode linear system of equations, which is exponentially unstable (see comments in chapter $5$ of \cite{TrefethenTextbook}) and nevertheless used in popular software packages in MATLAB and Python.\footnote{Vandermode matrices and their systems of linear equations are part of the implementation of methods such as polyfit in MATLAB and polyfit and chebfit in SciPy). }

In the case of Chebyshev interpolants, the optimal way to evaluate them is with the Barycentric interpolation formula (\cite{barycentricFormulaTrefethen}, \cite{barycentricFormula}).

\begin{theorem}\label{thm: barycentric formula}
Let $x_0, \ldots, x_n$ be a grid of Chebyshev points and let $v_0, \ldots, v_n$ be values associated to this grid. Then the Chebyshev interpolant associated to these points is given by

\begin{equation}\label{eq: barycentric formula}
p_n(x) =  \sum\limits_{i=0}^{n}\textsc{\char13}\frac{(-1)^{i}v_i}{x-x_i} \Bigg/  \sum\limits_{i=0}^{n}\textsc{\char13}\frac{(-1)^{i}}{x-x_i} 
\end{equation}

\noindent for values of $x$ not on the grid. For the special case when $x = x_i$, then $p(x) = v_i$. The primes on the summation mean that when $i = 0$ or $i = n$, then the expression is multiplied by $0.5$.
\end{theorem}

\begin{remark}\label{rmk: barycentric props}
There are several advantages to using Equation \ref{eq: barycentric formula}. The first is that only the values of the function $f$ at Chebyshev points are needed to evaluate $p_n(x)$. This means all that is needed to evaluate the interpolant is the tensor; there is no extra step required to go from Chebyshev tensor to Chebyshev interpolant. The second is that evaluating such formula requires linear effort $O(n)$ with respect to the degree of the polynomial. Thirdly, this formula is proven to be stable in floating point arithmetic for all $x$ within the domain of approximation \cite{BarycentricStability}. Moreover, it is scale-invariant, meaning that the formula does not change when we consider a general interval of the form $[a,b]$. The combination of Theorem \ref{thm: exponential convergence} and Theorem \ref{thm: barycentric formula} yield a technique that approximates functions to a high degree of accuracy by calling it a small number of times, where the resulting approximator, a polynomial of low degree, can be evaluated in no time at all in a numerically stable manner (see \cite{TrefethenTextbook} for more details).
\end{remark}

\begin{remark}\label{rmk: barycentric speed one dim}
To give an idea of the speed of the barycentric interpolation formula within the context of pricing function approximation, a degree $9$ polynomial ($10$ Chebyshev points), which would give a high level of accuracy for most pricing functions in finance due to Theorem \ref{thm: exponential convergence}, takes around $100$ nanoseconds per evaluation on a standard computer  using a single core.\footnote{The implementation for which the time was measured was done in C++.} If we are dealing with a risk calculation where $1,000$ evaluations need to be done, this would take $100$ microseconds or equivalently $0.0001$ seconds. A typical Monte Carlo simulation, where the number of evaluations is in the order of a million, it takes $0.1$ seconds.
\end{remark}

\subsection{Multi-dimensional case}\label{sec: cheb interpolants multi dim}

In this section we present extensions to higher dimensions of the concepts and results presented in Section \ref{sec: cheb interpolants dim one}. This is very important as most functions in finance are multi-dimensional.

\begin{definition}\label{dfn: chebpts multi}
Let $A$ be a hyper-rectangle in $\mathbb{R}^n$. That is, $A$ is defined as the Cartesian product of one-dimensional closed and bounded intervals $I_i$, $A = I_1\times \cdots\times I_n$. Let $\chi_i$ be Chebyshev points corresponding to the interval $I_i$, for all $i$, $1\leq i\leq n$. Let the number of Chebyshev points in $\chi_i$ be $m_i$. We define the grid of Chebyshev points on $A$ generated by $\chi_1, \ldots, \chi_n$ as the Cartesian product of the sets $\chi_i$, $\chi = \chi_1,\times\cdots\times \chi_n$. 
\end{definition}

\begin{figure}[H]
\centering
\includegraphics[width=8cm, height=9cm]{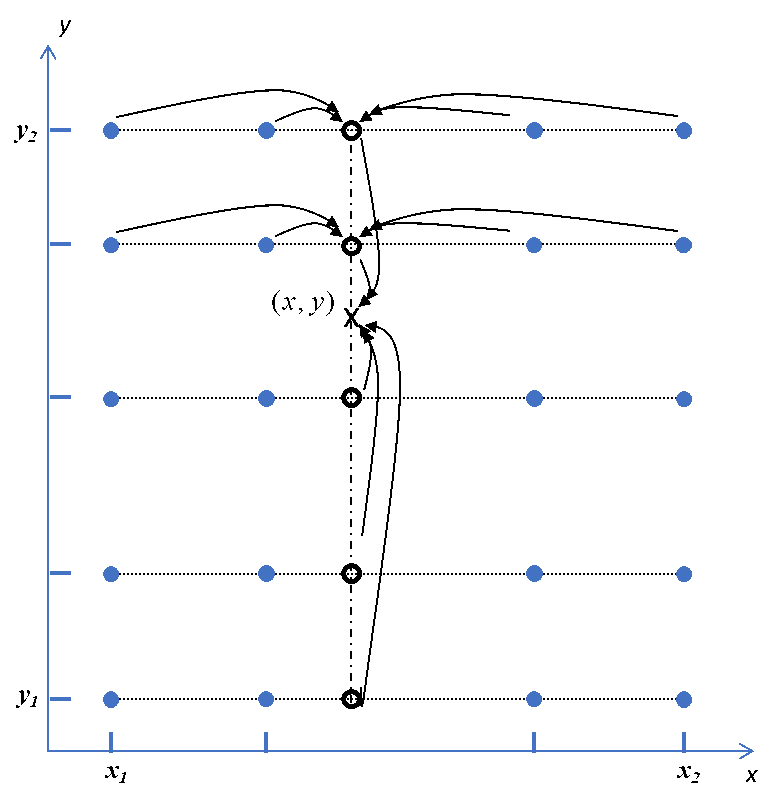}
\caption{Chebyshev grid in two dimensions.}
\label{fig: 2d cheb mesh}
\end{figure}

Using the notation in Definition \ref{dfn: chebpts multi}, the number of points on the multi-dimensional Chebyshev grid is $m_1\cdots m_n$. Figure \ref{fig: 2d cheb mesh} shows an example of a two-dimensional mesh. 

Say we have a $2$-dimensional function $f$ defined on $A$. Just as with the one-dimensional case, once the function has been evaluated on the mesh of Chebyshev points, a two-dimensional Chebyshev interpolant $p_{n,m}(x, y)$ is defined and ready to be evaluated. There are a number of proposed approaches to evaluate multidimensional Chebyshev frameworks (see \cite{Trefethen3D}, \cite{Trefethen2D} and \cite{GlauParamOptPric}). We have found the following to be optimal within practical settings. 

Without loss of generality, consider the point $(x, y)$. To evaluate the two-dimensional Chebyshev interpolant $p_{n,m}$ on $(x, y)$, consider the horizontal one-dimensional Chebyshev interpolants in Figure \ref{fig: 2d cheb mesh} and evaluate them at $x$. This gives values on the black circles of Figure \ref{fig: 2d cheb mesh}. These black circles lie on the horizontal lines defined by the Chebyshev points on the $y$-axis. Hence, the black circles, along with the values on them obtained from the evaluation of the horizontal one-dimensional Chebyshev interpolants, define another one-dimensional Chebyshev interpolant (running vertically as a dashed line in Figure \ref{fig: 2d cheb mesh}) that can be evaluated on $y$. The result of the latter evaluation is the value of $p_{n,m}$ at $(x, y)$.

The evaluation just described can be extended in a straightforward manner to higher dimensions. If we start with a Chebyshev mesh of dimension $n$, we evaluate a collection of one-dimensional Chebyshev interpolants to reduce the problem from $n$ dimensions down to $n-1$ dimensions. Continuing this way, the problem is reduced to the dimension one case, where the evaluation of the resulting one-dimensional Chebyshev interpolant gives the result.

To put the time this takes to run into context, let us make an estimate based on the time taken for a $1$-dimensional Chebyshev interpolant (see Remark \ref{rmk: barycentric speed one dim}). Assume a $3$-dimensional Chebyshev interpolant. Moreover, assume $10$ Chebyshev points per dimension. This gives a total of $1000$ Chebyshev nodes on the whole mesh. Given the evaluation algorithm described above, the barycentric interpolation formula is called 111 times which gives, assuming $100$ nanoseconds per barycentric interpolation formula call, $10$ microseconds per $3$-dimensional Chebyshev interpolation evaluation. If there are $1000$ scenarios to evaluate in a risk calculation, this would roughly take 10 milliseconds or $0.01$ seconds for the whole calculation. One million evaluations, such as the ones needed in a typical Monte Carlo simulation, takes $10$ seconds.

The following Theorem (\cite{GlauParamOptPric}) is the extension to higher dimensions of Theorem \ref{thm: exponential convergence}. Just as in the case of dimension one, when the function $f$ is analytic, the convergence of Chebyshev interpolants is as good as can be expected.

\begin{theorem}\label{thm: exponential multidim}
Let $f$ be a $d$-dimensional analytic function defined on $[-1,1]^d$. Consider its analytical continuation to a generalised Bernstein ellipse $E_p$, where it satisfies $\|f\|_{\infty}  \leq M$, for some $M$. Then, there exists a constant $C>0$, such that

\begin{equation*}
\|f-p_n \|_{\infty}\leq C\rho^{-m} 
\end{equation*} 

\noindent where $\rho=min_{(1\leq i\leq d)} \rho_i$, and $m=min_{(1\leq i\leq d)} m_i$. The collection of values $\rho_i$ define the radius of the generalised Bernstein ellipse $E_p$, and the values $m_i$ define the size of the Chebyshev mesh (see Definition \ref{dfn: chebpts multi}). For more details on Theorem \ref{thm: exponential multidim}, its proof and related results, see \cite{GlauParamOptPric}.
\end{theorem}

The combination of Theorem \ref{thm: exponential convergence}, Theorem \ref{thm: barycentric formula} and Theorem \ref{thm: exponential multidim} justifies the use of Chebyshev tensors and associated interpolants as a tool to approximate functions in finance.

\begin{remark}\label{rmk: not affected curse of dim}
An important note to make is that the application we are interested in, namely the computation of Dynamic Initial Margin, is not affected by the curse of dimensionality. This is a direct result of the methodology; that is, the way Chebyshev tensors are applied. We refer to Section \ref{sec: Sensitivities with Cheb} for details.
\end{remark}

\section{Computing sensitivities with Chebyshev tensors}\label{sec: Sensitivities with Cheb}

Simulating future IM involves computing IM at every node of a Monte Carlo simulation. In this section we present two ways in which Chebyshev tensors can be used to simulate sensitivities in an efficient and accurate manner. These sensitivities are then sued to compute DIM. The first method makes use of the risk factor evolution model, the second is agnostic to the model used. Note that the version of IM we compute is SIMM, the one proposed by ISDA and by now a standard in the industry for uncleared derivative transactions.

Consider a risk factor evolution model (RFEM) used to generate risk factors in a Monte Carlo simulation. Denote the dimension of the \emph{model space} by $k$. These are usually the number of stochastic factors in the model. For example, in the one-factor Hull and White (HW) model, this space consists of the short rate and hence $k$ = 1; a two-factor HW model has a model space with $k = 2$. In the context of Monte Carlo simulations for XVA or IMM, $k$ tends to be small.

Once the parameters of the RFEM have been calibrated they remain fixed throughout the simulation. At every node of the simulation, a set of values of the model space variables (short rate $r$, for example) fully determine the ISDA risk factors. These may include a multitude of interest rates, spreads, volatilities, etcetera. We call the space of ISDA risk factors the \emph{market space}. The latter typically has high dimension; sometimes in the hundreds. Denote the dimension of the market space by $n$.

Denote the function that generates ISDA risk factors from model space variables by $g$
 

\begingroup
\Large
\begin{equation}\label{eq: g function model space}
\begin{tikzcd}
\stackrel{  \textbf{ Model Space } } { \mathbb{R}^k}  \rar{g}  &  \stackrel{  \textbf{ Market Space  } } { \mathbb{R}^n}  
\end{tikzcd}
\end{equation} 
\endgroup


Functions like the one above are often analytic and lend themselves very well to be approximated with Chebyshev tensors. 

The following subsections describe the two ways in which dynamic sensitivities can be estimated using Chebyshev tensors. For illustration purposes we consider an Interest Rate Swap. Whatever is said about this example equally applies to any other trade type.

\subsection*{Model Space case}

Let the pricing function of the Swap be $f$. Computing SIMM requires computing the derivative of the pricing function $f$ with respect to each of the ISDA risk factors. This is difficult to obtain with a single tensor or interpolant due to the high dimension of $f$. Therefore, the dimension of the problem has to be reduced. The following, which makes use of the risk factor evolution mode, is one way. 

Consider a single time point within the Monte Carlo simulation. We need to compute the partial derivative of $f$ with respect to each ISDA risk factor. As an example, let the $i$-th swap rate $s_i$. Consider the following function $\varphi$

\begingroup
\Large
\begin{equation*}
\begin{tikzcd}
\mathbb{R}^k\rar{g}\arrow[black, bend right]{rr}[black,swap]{\varphi}  & \mathbb{R}^n \rar{S_i}  & \mathbb{R}
\end{tikzcd}
\end{equation*}
\endgroup

\noindent where $S_i$ denotes the partial derivative of $f$ with respect to $s_i$ 
\begin{equation*}
S_i = \frac{\partial f}{\partial s_i}
\end{equation*}

The $k$-dimensional function $\varphi$ is the result of composing two analytic functions. Therefore, it can be approximated very well using Chebyshev tensors. Note the function $\varphi$, given how it is defined, within the Monte Carlo simulation, gives the value of the partial derivative of $f$ with respect to $s_i$ at each node of the simulation.

To build a Chebyshev tensor for $\varphi$ do the following. Take the minimum and maximum value of each of the model space variables at the time point in question of the Monte Carlo simulation. For example, in the case of the Hull and White one-factor model, this would consist of the minimum and maximum values of the short rate at the time point in question of the Monte Carlo simulation. These values determine the hyper-rectangle (one-dimensional interval, in the case of Hull and White one-factor model) to which $\varphi$ is restricted. Notice, the hyper-rectangle just mentioned is contained in $\mathbb{R}^k$. Next, build a Chebyshev grid on this hyper-rectangle. Finally, call the function $\varphi$ on the Chebyshev grid. 

This is all the information needed to generate a Chebyshev tensor. Moreover, $\varphi$ is analytic (or smooth) as it is the composition of two analytic (smooth functions). Therefore, only a few Chebyshev points per dimension are needed due to Theorem \ref{thm: exponential multidim}. In the author's experience, between $4$ and $7$ points per dimension suffice for most applications.

The steps described above can be applied to every ISDA risk factor and every time point of the Monte Carlo simulation. This gives the whole distribution of ISDA sensitivities needed for the computation of SIMM. In Section \ref{sec: Results} we present the accuracy and computational gains obtained when this technique is applied to Swaps and Vanilla Swaptions.

\subsection*{Market Space case}

There is an alternative method that removes the dependency on the risk factor evolution model. The $k$-dimensional functions to be approximated using Chebyshev tensors are defined as follows. Consider the $i$-th swap rate $s_i$. The main challenge is to define a function $h$

\begingroup
\Large
\begin{equation}\label{Eq: h function Market Space}
h: \mathbb{R}^k \longrightarrow	\mathbb{R}^n
\end{equation}
\endgroup

\noindent that has the following condition. The function $h$ takes values in $\mathbf{R}^k$ and returns them in $\mathbf{R}^n$. As such, think of it as returning swap rate curves. The main condition demanded, is that the image of $h$ should contain all swap rate curves given by the simulation at the time point in question. That is, if the swap rate curve $(s_1, \ldots, s_n)$ has been generated by the Monte Carlo simulation, then there is $x$ in $\mathbb{R}^k$, such that $h(x)= (s_1, \ldots, s_n)$. Once a function $h$, as smooth as possible, with this characteristic, has been defined, the rest follows as before. That is, the following function is defined

\begingroup
\Large
\begin{equation*}
\begin{tikzcd}
\mathbb{R}^k\rar{h}\arrow[black, bend right]{rr}[black,swap]{\varphi}  & \mathbb{R}^n \rar{S_i}  & \mathbb{R}
\end{tikzcd}
\end{equation*}
\endgroup

\noindent and a Chebyshev tensor is built to approximate it. Essentially, the function $h$ plays the role of $g$ from the previous method, where $g$ is given by the risk factor evolution model.

There are several ways of defining $h$. The one that is described next was used to obtain the results presented in Section \ref{sec: Results}. Notice that it defines a function $h$ of one-dimension, keeping the Chebyshev tensor building time to a minimum.

Let $S_i$ represent the space spanned by the $i$-th swap rate $s_i$. Without loss of generality fix a time point in the simulation. Let $\{a_1, \ldots, a_m \}$ denote the values of the $i$-th swap rate generated by the Monte Carlo simulation at the time point in question.\footnote{Note this means there are $m$ paths in this Monte Carlo simulation.} Let $s$ be an arbitrary value of $s_i$ for which we need $h(s)$. We are only interested in finding values of the partial of $f$ with respect to $s_i$ at the nodes of the simulation. Therefore, $h$ only needs to be defined at values $s$ that lie in the interval defined by the maximum and the minimum of the set $\{a_1, \ldots, a_m \}$.


Consider $\alpha_1$ the element of $\{a_1, \ldots, a_m \}$, defined as the greatest of all values $a_i$ such that $a_i\leq s$. Similarly, consider $\alpha_2$ the smallest of all values $a_i$ such that $s\leq a_i$. Note that both $\alpha_1$ and $\alpha_2$ correspond to values of $s_i$ from the simulation and that $\alpha_1\leq s\leq \alpha_2$.

The value $h(s)$ that needs to be specified is an element of $\mathbb{R}^n$. Denote its $j$-th entry by $h(s)_j$. The value $h(s)_i$, is simply given by $s$. This is the case as we assume to be defining $h$ for the $i$-th swap rate. For the remaining entries, that is, all those $j\neq i$, where $1\leq j\leq n$, do the following. Take the $j$-th swap rate $s_j$. Given that $\alpha_1$ and $\alpha_2$  are values from the simulation, there are values $\beta_1$ and $\beta_2$ in the space spanned by $s_j$, that correspond to the same swap rate curves of $\alpha_1$ and $\alpha_2$, respectively. Finally, interpolate between $\beta_1$ and $\beta_2$ to obtain $h(s)_j$.

If the IM calculation requires $l$ sensitivities, this method creates $l$ Chebyshev tensors of dimension one per time point. Each of these Chebyshev tensors is evaluated at each node of the time step ($m$ of them) of the Monte Carlo simulation to obtain the sensitivities needed for the computation of SIMM.


Section \ref{sec: Results} presents the results of the simulations run with the Chebyshev techniques just presented. In Section \ref{sec: Results analysis} we analyse the results and touch on the advantages and disadvantages of using one over the other, along with further comparisons to other non-Chebyshev based techniques.

\section{Results}\label{sec: Results}

The Monte Carlo simulation used to produce results consisted of $10,000$ paths. Different number of time points in the future were used depending on the trade type.\footnote{Running brute force simulations for DIM is very expensive. To run tests within a reasonable time different number of time points were used for Swaps compared to Swaptions.} Two types of trades were chosen: Interest Rate Swaps and European Swaptions. The interest rate curves were simulated using a one-factor Hull-White model while the volatility was simulated using a one-factor SABR model.

Different methodologies were used to compute DIM. The benchmark was obtained using the original pricing function to compute the partial derivatives at each node of the simulation using finite difference in a “brute-force” fashion. The alternative methodologies, all compared to the benchmark in terms of speed and accuracy, consist of the two Chebyshev techniques described in Section \ref{sec: Sensitivities with Cheb}, and the two regression techniques described in \cite{Zhu Chan}. The first regression technique is polynomial regression, the second is the Nadaraya-Watson kernel regression. For more details on how to implement these regressions techniques for the computation of DIM, we refer to \cite{Zhu Chan}.

\begin{figure}
\centering
\includegraphics[scale=0.15]{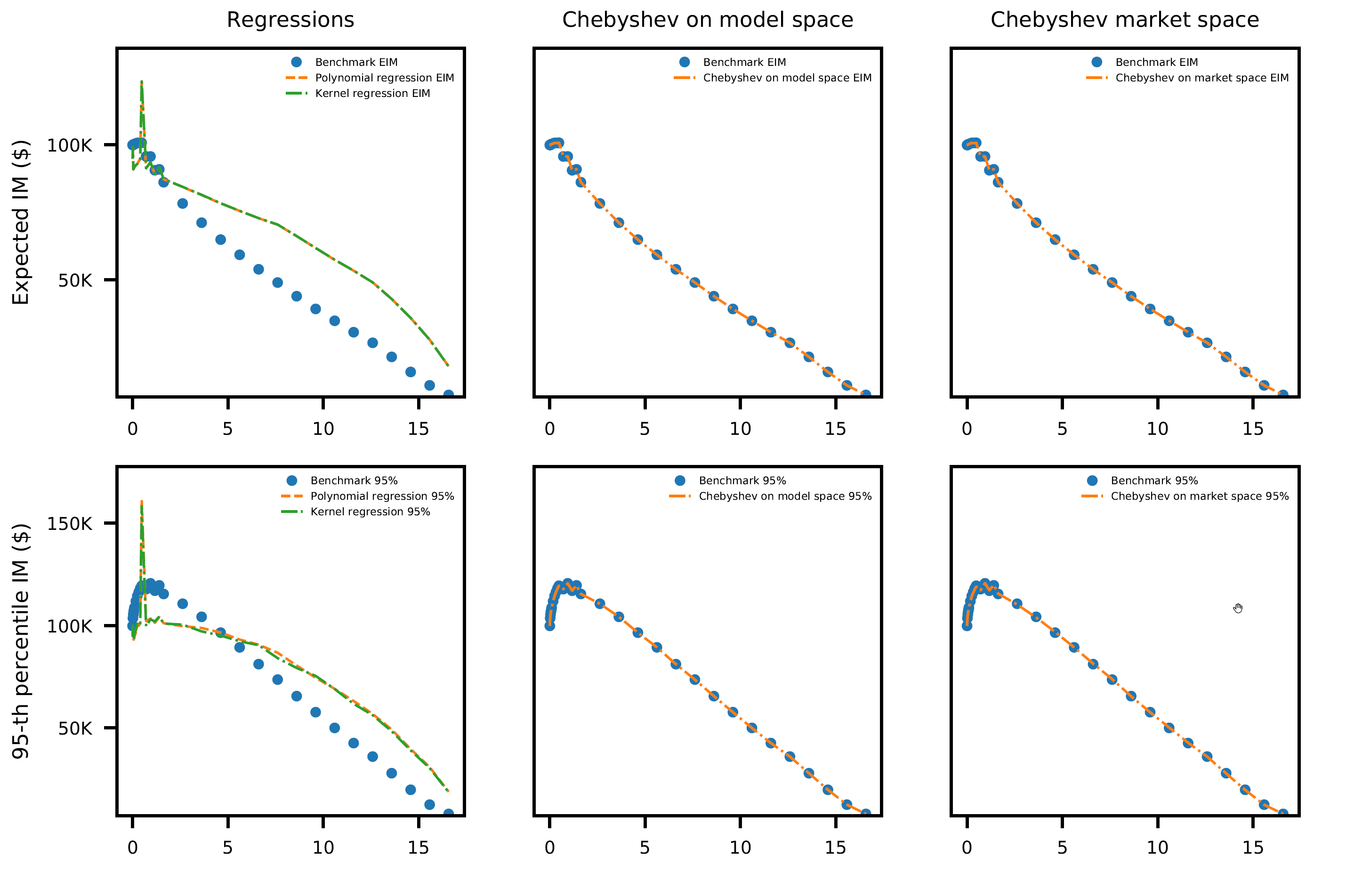}
\caption{DIM profiles for the Interest Rate Swap. Comparison between brute force and regressions (first column), and brute force with Chebyshev techniques (second and third column).}
\label{fig: Swap}
\end{figure}

Figure \ref{fig: Swap} shows comparisons between the Expected Profile of IM (EIM) and the 95-th percentile of IM throughout the simulation for all methodologies used for a single Interest Rate Swap. The Monte Carlo simulation consisted of $10,000$ paths and $30$ time points giving a total of $300,000$ simulation nodes. At each node sensitivities to all risk factors are needed for the computation of SIMM. 

The first row in Figure \ref{fig: Swap} shows a comparison of the EIM between the brute force approach and one of the techniques: regressions and the two different ways of applying Chebyshev tensors. The second row presents the corresponding comparisons for the $95$-th quantile profile.

Table \ref{tab: Errors Cheb swap} contains the relative errors, expressed as percentages, between the benchmark and each of the alternative methodologies, both for EIM and the $95$-th percentile profile. The metric is computed as follows. Denote the benchmark profile by \emph{pbm} and the profile obtained by any given alternative methodology by \emph{palt}. Then the error metric is

\begin{equation}\label{eq: error metric}
\text{mean}\bigg( \frac{\text{abs}( \text{pbm} - \text{palt})}{ \text{pbm}} \bigg)
\end{equation}

That is, the absolute value of the difference (normalised by the benchmark value) is first obtained at each time point. Then the average across time step gives the metric.

\begin{table}
\centering
\begin{tabular}{|l|l|l|}
\hline
DIM profile & EIM & 95$\%$ \\
\hline
Chebyshev on Model Space &	$0.0000003$ & $0.0000006$ \\
\hline
Chebyshev on Market Space &	$0.01$ & $0.5$ \\
\hline
\end{tabular}
\caption{Accuracy of Chebyshev techniques applied to the computation of SIMM profiles for an Interest Rate Swap. The profiles computed are expected DIM exposure and the $95\%$ quantile.}
\label{tab: Errors Cheb swap}
\end{table}

\begin{figure}[H]
\centering
\includegraphics[scale=0.45]{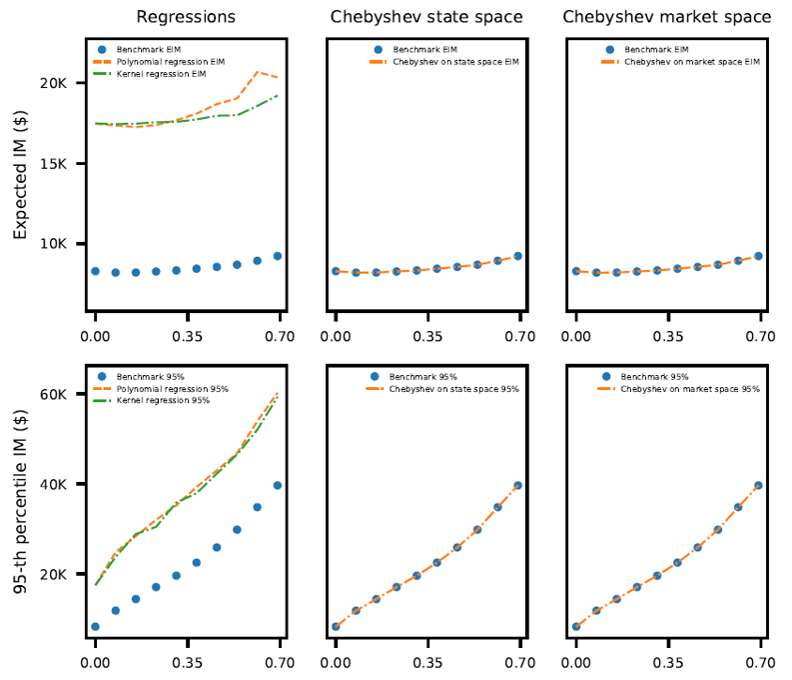}
\caption{DIM profiles for at the money Swaption. Comparison between brute force and regressions (first column), and brute force with Chebyshev techniques (second and third column).}
\label{fig: Swaption ATM}
\end{figure}

\begin{figure}
\centering
\includegraphics[scale=0.45]{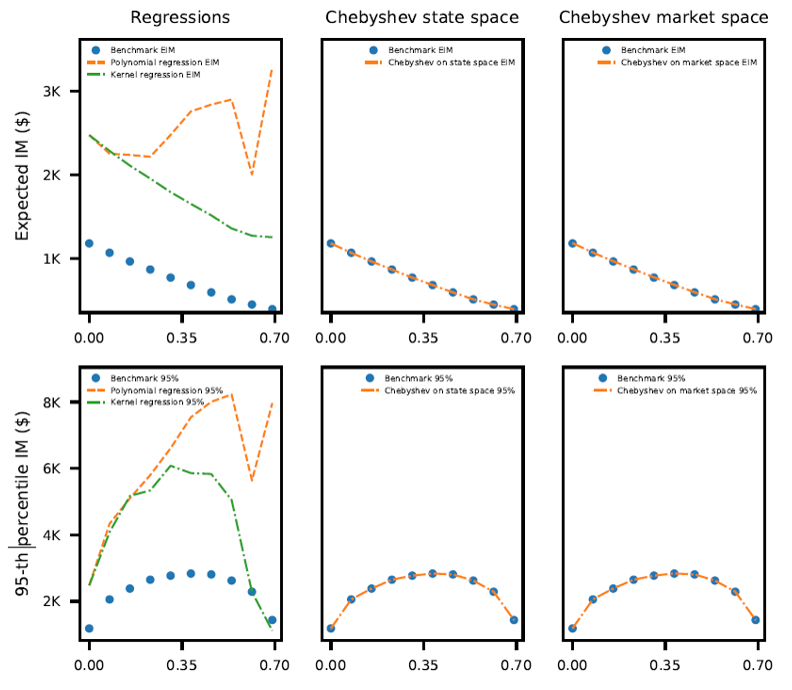}
\caption{DIM profiles for out of the money Swaption. Comparison between brute force and regressions (first column), and brute force with Chebyshev techniques (second and third column).}
\label{fig: Swaption OTM}
\end{figure}

\begin{figure}
\centering
\includegraphics[scale=0.45]{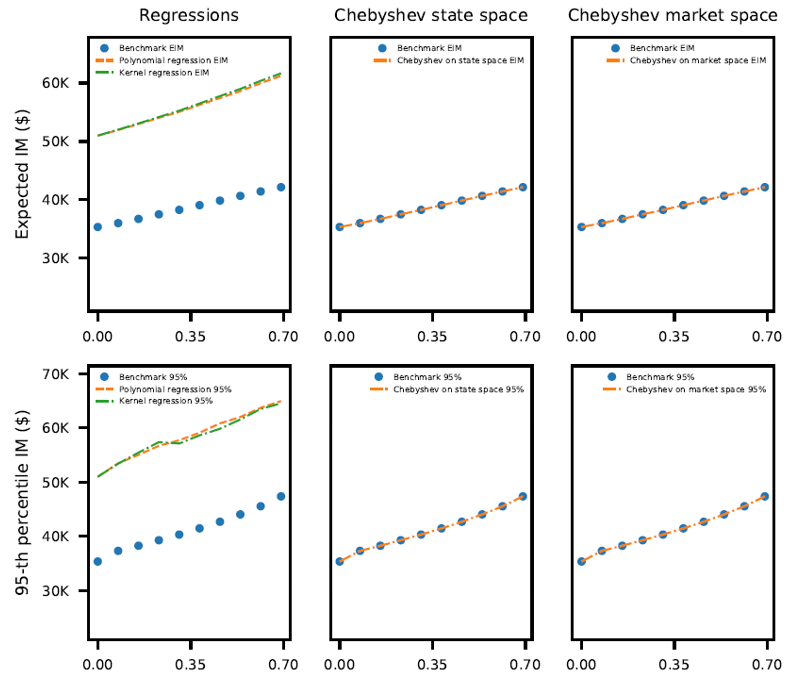}
\caption{DIM profiles for in the money Swaption. Comparison between brute force and regressions (first column), and brute force with Chebyshev techniques (second and third column).}
\label{fig: Swaption ITM}
\end{figure}

Figures \ref{fig: Swaption ATM}, \ref{fig: Swaption OTM} and \ref{fig: Swaption ITM} show the corresponding results for an out-of-the-money (OTM), at-the-money (ATM) and in-the-money (ITM) Swaption, respectively. The maturity of each swaption was chosen to be short to increase curvature and fully test the different methods. The Monte Carlo simulation used for Swaptions consisted of $10,000$ paths and $12$ time points in the future giving a total of $120,000$ simulation nodes.

\bigskip

\bigskip

Table \ref{tab: Errors Cheb swaptions} shows the errors for each of the Swaptions and for each of the approximation techniques used, obtained using Equation \ref{eq: error metric}.
\bigskip

\begin{table}[H]
\centering
\begin{tabular}{|l|l|l|l|l|l|l|}
\hline
Swaption moneyness & \multicolumn{2}{|c|}{ATM} & \multicolumn{2}{|c|}{OTM} & \multicolumn{2}{|c|}{ITM} \\
\hline
DIM profile & EIM & 95$\%$ & EIM & 95$\%$ & EIM & 95$\%$\\
\hline
Chebyshev on Model Space &	$0.00003$ & $0.0005$ & $0.0002$ & $	0.002$ & $0.000006$ & $0.0001$\\
\hline
Chebyshev on Market Space &	$0.004$ & $0.006$ & $0.3$ & $	0.08$ & $0.0007$ & $0.001$\\
\hline
\end{tabular}
\caption{Accuracy of Chebyshev techniques applied to the computation of SIMM profiles for Swaptions with short maturity and three different moneyness (At the money, Out of the money, In the money). The profiles computed are expected DIM exposure and the $95\%$ quantile.}
\label{tab: Errors Cheb swaptions}
\end{table}


\section{Discussion of Results}\label{sec: Results analysis}

As mentioned in Section \ref{sec: Intro}, the first requirement of a good risk metric approximation technique is accuracy. Clearly, both Chebyshev techniques are orders of magnitude better than regression techniques. Moreover, as Chebyshev techniques give a very accurate replica of the brute-force calculation, they should capture the sensitivities of MVA to initial market conditions allowing for accurate hedging. In addition to this, the Chebyshev techniques presented in this paper approximate the sensitivities of each trade with respect to its risk factors. This means that as a side consequence, we end up with dynamic sensitivities, which can be used for dynamic hedging.

Notice the accuracy obtained when building Chebyshev tensors on the Model Space is orders of magnitude higher than the one obtained when building them on the Market Space. This is a direct consequence of the high degree of smoothness of function $g$ (Equation \ref{eq: g function model space}). As $g$ and $S_i$ are smooth, their composition $\varphi$, for which a Chebyshev tensor is built, is approximated to a very high degree of accuracy with only a few grid points. The downside of this approach is that the dimension of the Chebyshev tensors is the dimension $k$ of the RFEM used. Therefore, the build time of the Chebyshev tensors increases as $k$ increases.

On the other hand, the Market Space approach defines one dimensional functions $\varphi$ to be approximated by Chebyshev tensors independent of the value $k$, keeping build times low. This, however, comes with a cost in accuracy. The latter a consequence of the fact that the function $h$ (Equation \ref{Eq: h function Market Space}) is not guaranteed to be smooth. The loss in accuracy can be seen in Tables \ref{tab: Errors Cheb swap} and \ref{tab: Errors Cheb swaptions}, where the accuracy of the Model Space approach is orders of magnitude greater than the Market Space approach. Notice, however, that even the latter offers, for the cases studied in this paper, very high levels of accuracy.

The regression techniques used in this paper approximate IM distributions where IM is defined as the quantile of PnLs distributions. Strictly speaking, the results obtained with regressions should be compared to a brute force simulation of DIM where IM is computed via quantiles. Although no such thing was done in \cite{Zhu Chan} (possibly due to the computational demand behind this calculation), statistical tests to assess the consistency of the technique were done. These tests showed some good results but also some shortcomings due to the number of assumptions made, such as the normality of the PnL distributions. For more details see \cite{Zhu Chan}.

To compute profiles of Dynamic SIMM using regressions, the authors of \cite{Zhu Chan} suggest shifting the profiles obtained with regressions by the scalar value needed to match their definition of IM  (via quantiles) with SIMM at $t_0$ in the Monte Carlo simulation. Although this is common practice by some in the industry, it is an approach which is often frowned upon by regulators. Part of the big differences between the IM profiles obtained with regressions and those obtained through brute force and Chebyshev (in Figures \ref{fig: Swaption ATM}, \ref{fig: Swaption OTM} and \ref{fig: Swaption ITM}), can be explained by the fact that they estimate IM using different definitions. Although applying the shifting factor suggested in \cite{Zhu Chan} would improve results, specially close to $t_0$, we decided against it as this would not be a practice allowed in production systems. Figures \ref{fig: Swaption ATM}, \ref{fig: Swaption OTM} and \ref{fig: Swaption ITM} show that regression techniques are not good estimators of Dynamic SIMM. This should come as no surprise as SIMM is driven by sensitivities while regression techniques disregard this aspect of the calculation.

There are two popular techniques used in the computation of DIM which were not tested in this paper. These are Adjoint Algorithmic Differentiation (AAD) and Deep Neural Nets. For the sake of completeness, we comment on the accuracy normally reported in the literature. AAD is a well known technique that has the major advantage of computing sensitivities to a very high degree of accuracy. This in turn leads to very accurate DIM values. Deep Neural Nets have recently been applied with success in the computation of DIM (for example, see \cite{DNNs IM}). In \cite{DNNs IM}, relatively simple Deep Neural Nets were trained with relatively few training samples, giving average errors in the range of $1\%$ to $3\%$.

%
%

To asses the computational burden of each of the approaches tested in this paper, we break it up in three components. The first is the cost of calling the pricing function. This is the most expensive component in each technique. It is what is referred to in Table \ref{tab: Comp burden and accuracy} as pre-compute effort. This includes, for example, the cost of computing Initial Margin at each node of the simulation (using pricing functions), in the case of regressions, or evaluating sensitivities to the trades at each node, in the case of Chebyshev tensors. The second component comes in the form of training; for example the training of the Deep Neural Nets. In most cases, this component is not significant. The smallest of all three components comes in the form of evaluation. A common characteristic of all techniques is that once the pre-computation and training has been done, the evaluation of the approximating objects (regression functions, Deep Neural Nets, Chebyshev tensors) is very fast. Notice as well, that computing SIMM at each node of the simulation, once sensitivities are available, require calculations that run very efficiently on a computer, hence we ignore this part of the calculation.


Let us quantify what was just described. Consider a Monte Carlo simulation with  $10,000$ paths and $100$ time steps. This simulation has $1,000,000$ nodes and hence, if we were to price a portfolio at ever node, this would have $O(10^6)$ pricing cost. This represents the cost of a typical CVA calculation. Regression techniques, as used in \cite{Zhu Chan} rely on having PV values at each node of the simulation and at ten days ahead of each node. This gives $2,000,000$ calls to the pricing functions which has $O(10^6)$ cost. The Chebyshev methods presented in this paper rely on building Chebyshev tensors for each sensitivity at each time point. Assuming between $10$-$50$ sensitivities per trade (notice these are non-zero sensitivities), $10$ Chebyshev nodes per tensor (when tensors are built on the market space, giving one-dimensional tenors) and $100$ time points, we have $O(10^4)$ calls to the pricing function. That is, two whole orders of magnitude less than regressions. If tensors are built on the model space, we build low dimensional tensors, say with $3$ dimensions, which typically require around $100$-$200$ points. This gives $O(10^5)$ calls to the pricing function, which is still an order of magnitude less than regressions and a typical CVA calculation.

The other two popular methods mentioned above, not tested in this paper, require the following computational burdens. AAD typically incurs in $5$ to $10$ times the cost of a CVA calculation. That is, $O(10^6)$ to $O(10^7)$ pricing cost. Deep Neural Nets, although fast upon evaluation, also require, just like regressions, the PVs at every node of the simulation (\cite{DNNs IM}). That is $O(10^6)$ calls to the pricing functions. In all cases considered, Chebyshev tensors, applied either on the Model Space or the Market Space, incur in at least a whole order of magnitude less. The previous comments are summarised in table \ref{tab: Comp burden and accuracy}.



\begin{table}[H]
\centering
\renewcommand{\arraystretch}{1.35} 
{\footnotesize
\begin{tabular}{|c |c|c|c|c| c|}
\hline
 & \thead{\textbf{Pre-compute effort} \\ \textbf{(off-line)}} & \thead{\textbf{Training} \\ \textbf{effort}} & \thead{\textbf{Evaluation} \\ \textbf{effort}} & \thead{\textbf{Total} \\ \textbf{effort}} &  \thead{\textbf{Accuracy}} \\
\hline
Brute force & $O(10^7)$ & n/a & $\times 10$ &  n/a & Benchmark\\
\hline
Chebyshev & $O(10^4)\sim O(10^5)$ &	 n/a  & $\sim 0$ & $O(10^4)\sim O(10^5)$ & Very high \\
\hline
DNNs & $O(10^6)$ &	$+ 10^3$  & $\sim 0$ & $O(10^6)$ & Medium \\
\hline
Regressions & $O(10^6)$ &	$\sim 0$  & $\sim 0$ & $O(10^6)$ & low \\
\hline
AAD & $O(10^6)\sim O(10^7)$ &	n/a  & n/a & $O(10^6)\sim O(10^7)$ & Very high \\
\hline
\end{tabular}
}

\caption{Computational burden and accuracy in the computation of DIM for each of the different techniques considered. The computational burden is broken up into pre-compute, training and evaluation efforts. Notice that under the assumptions made about a typical Monte Carlo simulation in this section, a CVA calculation costs $O(10^6)$.}
\label{tab: Comp burden and accuracy}
\end{table}

%
%

With respect to the third criteria presented in Section \ref{sec: Intro}, ease of implementation, regressions, Deep Neural Nets and Chebyshev interpolants can be implemented within an engine with minimal intrusion. In the case of Chebyhsev, all that is needed is the price of the trade or portfolio at a few critically-selected points. In the case of regressions and Deep Neural Nets, a whole distribution of PVs on the Monte Carlo simulation is required. Assuming this is available, the training of both regressions and DNNs is quick and relatively straightforward. 

There is an important observation to make with regards to the use of Deep Neural Nets. For the cases test in \cite{DNNs IM}, the architecture of the Deep Neural Nets used was relatively simple. However, the choice of this hyper-parameter (and others, such as activation function) is expected to increase in difficulty as the complexity and dimension of the trades increase.

All three objects, Chebyshev tensors, regressions and Deep Neural Nets are simple enough that can be stored in memory to be used in other intra-day calculations and for other calculations in the future. All this with small memory footprints.

Out of the four techniques mentioned in this paper, the one that stands out for its difficulty of implementation is AAD. Not only does it impose a substantial memory demand, but its implementation within existing engines can be extremely cumbersome.

\section{Conclusion}\label{sec: Conclusion}

Chebyshev tensor enjoy remarkable mathematical properties that make them ideal candidates to approximate analytic functions, such as pricing functions, to high degrees of accuracy with little computational effort (see Section \ref{sec: cheb tensors}). This paper shows how to harness the power of Chebyshev tensors to compute Dynamic Initial Margin (SIMM) to a high degree of accuracy, with low computational cost, and little implementation effort. 

Dynamic Initial Margin simulations were run on a Swap and Swaptions of different moneyness and short maturity (forcing it to exhibit more of its non-linearities). The techniques implemented and tested consisted of two Chebyshev techniques which compute dynamic sensitivities (Section \ref{sec: Sensitivities with Cheb}), and the two types of regression presented in \cite{Zhu Chan} that directly compute DIM, as defined through quantiles. The accuracy and speed of the different techniques considered were measured with respect to a benchmark. The benchmark considered was the calculation of DIM using pricing functions, such as the ones found in Front Office systems. 

The accuracy and speed obtained with both Chebyshev techniques are remarkable (Section \ref{sec: Results}). The regressions used in \cite{Zhu Chan} provide decent speed-ups but at the cost of low accuracy (see Section \ref{sec: Results analysis}). 

Further comparisons were also made with other techniques to compute DIM that have recently been reported in the literature; namely AAD and and Deep Neural Nets (as in \cite{DNNs IM}).

All techniques considered (whether implemented or not) have advantages and disadvantages as discussed in Section \ref{sec: Results analysis}. Overall, given the evidence, Chebyshev tensors provide the best balance of them all when measured in terms of accuracy, computational cost and ease of implementation.

The numerical results presented in this paper leave the door open to further tests. Specifically, using Risk Factor Evolution Models with a Model Space of greater dimension than the one considered in this paper. With respect to this, tt is important to note, however, that the description presented in Section  \ref{sec: Sensitivities with Cheb} on how to apply the Chebyshev tensors still applies in higher dimensions. It is only a matter of implementation and testing. Results, due to the mathematical properties described in \ref{sec: cheb tensors}, are expected to be very good too.

\pagebreak


\end{document}